\documentclass[12pt]{article}

\usepackage{amsmath, amssymb, amsthm}
\usepackage{booktabs}
\usepackage{multirow}
\usepackage{graphicx}
\usepackage{float,subcaption}
\usepackage{algorithm}
\usepackage{url}
\usepackage{xurl}
\usepackage{hyperref}
\usepackage[noend]{algpseudocode}
\usepackage[round,authoryear]{natbib}
\usepackage{geometry}
\title{Bayesian Reconciliation of Overlapping Multi Step
Forecasts: A Post Processing Framework for
Cross Origin Coherence}

\author{
    Oluwatosin Bamidele Fajana\thanks{University of S\~ao Paulo, Brazil \textit{and} Federal University of S\~ao Carlos, Brazil} \and 
    Ricardo S. Ehlers\thanks{University of S\~ao Paulo, Brazil} \and 
    Marinho G. Andrade\thanks{University of S\~ao Paulo, Brazil}
}

\date{}

\begin{document}

\maketitle

\begin{abstract}
Overlapping multi step forecasts often disagree, a form of vertical instability that we term cross origin incoherence. We propose a Bayesian reconciliation framework that treats overlapping forecasts as noisy, biased measurements of a common latent trajectory. Operating as a model agnostic post processing layer, it requires no access to model internals. Simulations demonstrate substantial incoherence reduction and robust latent trajectory recovery, remaining highly resilient (correlation 0.972) even under severe nonlinear stress tests. An empirical application to S\&P 500 volatility evaluates AR(1), ETS, and LSTM models. The framework reduces incoherence by over 97\% across all models while delivering fully calibrated probabilistic forecasts (93\% to 94\% coverage). Notably, it improves point accuracy for ETS and LSTM. Furthermore, the estimated horizon specific parameters act as a diagnostic X-ray, revealing hidden biases in neural networks without requiring access to their architecture.
\end{abstract}

\noindent\textbf{Keywords:} Bayesian reconciliation, vertical stability, multi step forecasting, cross origin, coherence, incoherence, state space models, probabilistic post processing, uncertainty quantification

\section{Introduction}
\label{sec:introduction}

Multi step forecasting systems, including modern AI models such as LSTM and transformer networks, can generate internally inconsistent forecasts. Suppose a volatility forecaster predicts on Monday that Friday's market volatility will be 0.02. On Tuesday, the same model, trained on the same data with the same architecture, predicts that Friday's volatility will be 0.03. Which forecast should a risk manager trust? Both cannot be correct, yet both come from the same model. This is not statistical noise; it is a violation of logical coherence. Forecasts for the same future event should not contradict each other depending on when they were made. We call this failure mode cross origin incoherence: disagreement among overlapping forecasts that target the same future outcome but are issued at different forecast origins. The framework proposed here is intentionally agnostic to model architecture. Its purpose is not to replace forecasting models but to correct a structural failure mode common to complex multi step prediction systems.

The degree of incoherence can be measured by how much forecasts for the same target time disagree. Let $\hat{y}_{t|i}$ denote the forecast for time $t$ made from origin $i$. The cross origin incoherence metric is
\begin{equation}
\mathcal{I} = \frac{1}{T}\sum_{t=1}^{T}\frac{1}{|\mathcal{S}_t|}\sum_{(i,j)\in\mathcal{S}_t}\left(\hat{y}_{t|i}-\hat{y}_{t|j}\right)^2,
\label{eq:incoherence}
\end{equation}
where $\mathcal{S}_t$ is the set of all distinct forecast pairs $(i,j)$ that both target time $t$. To see the structure of the metric, arrange the forecasts into the matrix
\begin{equation}
\hat{\mathbf{Y}} =
\begin{bmatrix}
\hat{y}_{1|1} & \hat{y}_{2|1} & \hat{y}_{3|1} & \cdots & \hat{y}_{T|1} \\
\cdot         & \hat{y}_{2|2} & \hat{y}_{3|2} & \cdots & \hat{y}_{T|2} \\
\cdot         & \cdot         & \hat{y}_{3|3} & \cdots & \hat{y}_{T|3} \\
\vdots        & \vdots        & \vdots        & \ddots & \vdots \\
\cdot         & \cdot         & \cdot         & \cdots & \hat{y}_{T|T}
\end{bmatrix},
\label{eq:forecast_matrix}
\end{equation}
whose rows index forecast origins $i$ and whose columns index target times $t$. The entry in row $i$ and column $t$ is $\hat{y}_{t|i}$, the forecast for time $t$ issued at origin $i$. The lower triangle is empty because forecasts are only issued for the current or future target times; the symbol $\cdot$ marks entries that are not issued. For a given target time $t$ with $n_t$ forecast origins, the set $\mathcal{S}_t$ contains $\binom{n_t}{2}=n_t(n_t-1)/2$ distinct pairs. The inner sum in \eqref{eq:incoherence} therefore gives the average squared disagreement among forecasts of target $t$, and the outer sum averages these column wise disagreements across the $T$ target times. Normalization by $|\mathcal{S}_t|$ ensures comparability across series lengths and horizons, and a perfectly coherent system satisfies $\mathcal{I}=0$.

Throughout, we use the terms cross origin incoherence, forecast contradiction, and logical inconsistency interchangeably. While the concept of measuring disagreement across forecast origins is related to the notion of forecast congruence \citep{pritularga2024forecast} and vertical stability \citep{godahewa2025stability}, the metric in Equation~\eqref{eq:incoherence} is operationally distinct in two ways. First, rather than computing the variance around a mean forecast, we calculate the average pairwise squared difference. This pairwise formulation provides a direct measure of the expected distance between any two forecast origins without requiring a central reference point. Second, the normalization by \(|\mathcal{S}_t|\) dynamically adjusts for the varying number of available forecast origins at each target time. This prevents the metric from being artificially deflated at the end of the forecast horizon, where the triangular forecast matrix yields fewer overlapping predictions. We use the squared formulation because it mirrors mean squared error and penalizes severe contradictions more heavily \citep{gneiting2007,timmermann2006}. The metric is used here specifically as a diagnostic tool to measure the degree to which the identity constraint of the latent state model is violated, not as a new stability measure in itself. The contribution of the paper lies elsewhere: in the Bayesian reconciliation framework that maps an already generated forecast matrix to a coherent latent trajectory.

The problem of reconciling conflicting forecasts spans four related literatures. \emph{Forecast combination.} Classical pooling treats overlapping predictions for a common target as distinct predictors within an information set \citep{bates1969combination,granger1984improved}. Optimal linear pooling and time varying weights use historical performance to merge these signals \citep{timmermann2006}. These methods optimize aggregate point error, but they do not guarantee that successive forecasts for the same target agree, and they treat each forecast trajectory as static rather than accounting for the decay of accuracy with horizon.

\emph{Hierarchical and cross temporal reconciliation.} The modern reconciliation paradigm of \citet{hyndman2011} and \citet{wickramasuriya2019} enforces coherence through linear constraints across aggregation structures, so that regional forecasts sum exactly to national totals. \citet{athanasopoulos2017} extended this idea to temporal hierarchies, and \citet{difonzo2023}, \citet{girolimetto2024}, \citet{athanasopoulos2024}, and \citet{girolimetto2026} developed cross temporal and probabilistic extensions. Minimum trace (MinT) projections guarantee linear coherence across aggregation levels, but they are not designed for the localized, origin specific bias trajectories that arise when a single model's predictions are updated across sequential forecast origins. Coherence in these methods is defined relative to an aggregation structure. The present paper addresses a different constraint: the identity constraint that all forecasts of a common target timestamp should agree with each other.

\emph{Deterministic stabilizers.} \citet{godahewa2025stability} formalize vertical forecast stability and propose deterministic post processing that smooths forecasts along diagonals of the forecast matrix, with an explicit stability accuracy trade off. Their method is efficient and model agnostic, but it operates on point forecasts and provides neither uncertainty quantification nor horizon specific reliability diagnostics. The meteorological literature addresses the same phenomenon under the name jumpiness \citep{zsoter2009jumpiness}, with lagged ensembles \citep{hoffman1983lagged} as the standard remedy. In short, \citet{godahewa2025stability} provide a deterministic linear stabilizer applicable to any base model; the present framework provides a probabilistic Bayesian alternative that enforces coherence while additionally quantifying uncertainty and recovering horizon specific reliability.

\emph{Stability aware training and calibration.} \citet{ma2026beyond} introduce a forecast accuracy and coherence score that is model agnostic as a metric and can be applied to any probabilistic multi horizon forecast. Their empirical instantiation optimizes this score as a differentiable training objective for SARIMA models, which requires a differentiable forecaster and access to the training pipeline. Deep neural forecasters such as DeepAR \citep{salinas2020} and LSTM based models are now standard in industrial applications \citep{benidis2020,lim2021}, but they are trained for accuracy horizon by horizon and provide no formal guarantee that successive forecasts for a common target agree. Calibration tools operate on individual intervals: Monte Carlo dropout \citep{gal2016}, deep ensembles \citep{lakshminarayanan2017}, and conformal prediction \citep{angelopoulos2023} each improve the calibration of a predictive interval, but none enforces agreement across origins. An AI model can predict a sharp rise tomorrow followed by a gentle decline and, on the next day, a gentle rise followed by a sharp decline, both with well calibrated uncertainty. Stability aware training penalizes this ex ante; the present framework enforces coherence ex post without retraining.

The closest methodological antecedent is the Bayesian state space reconciliation framework of \citet{eckert2021}, who model reconciliation errors as latent states under an aggregation constraint. Their latent state is a reconciliation error, and the constraint is hierarchical. The latent state in the present paper is the coherent trajectory itself, and the constraint is an identity constraint at each target time rather than a linear aggregation constraint. The horizon specific bias and reliability parameters estimated here have no counterpart in hierarchical reconciliation, and the two approaches are complementary.

We bridge these strands with a Bayesian state space framework for forecast reconciliation. The core idea is to reinterpret the forecaster's trajectory not as a collection of final predictions but as noisy measurements of a latent coherent process. By treating all forecasts that target a common future time, regardless of origin, as informing a single latent state, coherence is enforced by construction. Because the method operates exclusively on the forecaster's output matrix, it requires no access to the model's internal parameters, architecture, or training data; the observed series is used only for final evaluation. The rolling AR(1) benchmark in the empirical section is chosen for transparency, not because it is the intended application; it allows the effects of coherence enforcement to be isolated and interpreted.

The contribution is threefold. First, we formulate cross origin reconciliation as a latent state inference problem, distinct from constraint based reconciliation and from training time stability penalties. Second, we develop a Bayesian reconciliation framework that enforces coherence through a latent state representation while remaining entirely model agnostic, and we establish identifiability of the model parameters, ensuring that the reconciled trajectory is uniquely defined and interpretable. Third, we show through controlled simulation and an empirical application to S\&P 500 volatility that coherence can be substantially reduced without retraining or modifying the underlying forecasting system, and we quantify the accuracy calibration trade off that this entails. The contribution is methodological; it is not a claim that forecast stability or forecast reconciliation is itself a novel concept.

State space models and the Kalman filter \citep{durbin2012} provide the machinery for estimating latent trajectories from noisy observations. Traditionally these models smooth observed data to estimate an underlying state. We repurpose the machinery to reconcile predictions instead of smoothing observations, and this conceptual shift is central to the contribution. We implement the model in Stan \citep{carpenter2017} using Hamiltonian Monte Carlo with the No U Turn Sampler \citep{hoff-gelm}. To our knowledge, no existing work enforces cross origin coherence across the multi step forecast trajectory of a single black box forecasting model via a purely post processing, Bayesian state space approach. This claim distinguishes the specific methodological combination proposed here rather than suggesting that the problem of forecast stability is unexplored; \citet{ma2026beyond} and \citet{godahewa2025stability} address related problems from different directions.

The remainder of the paper is organized as follows. Section~\ref{sec:model} introduces the Bayesian state space framework and the posterior inference procedure. Section~\ref{sec:simulation} evaluates the method through controlled simulations designed to assess parameter recovery and coherence enforcement. Section~\ref{sec:empirical} presents the empirical application to S\&P 500 volatility. Section~\ref{sec:conclusion} discusses assumptions, limitations, and directions for future research.

\section{Model Framework}
\label{sec:model}

Consider a time series \(\{y_t\}_{t=1}^T\) that we wish to forecast. At each time \(t\), a pre-trained forecasting model produces a vector of \(H\) step ahead point forecasts:
\[
\hat{\mathbf{y}}_t^{\mathrm{AI}} = (\hat{y}_{t+1|t}, \hat{y}_{t+2|t}, \dots, \hat{y}_{t+H|t})^\top.
\]
We use \(\hat{y}_{t+h|t}\) and \(\hat{y}_{t|i}\) interchangeably to denote a forecast for target time \(t+h\) issued at origin \(t\). These forecasts are cross origin incoherent if there exists no underlying stochastic process \(\{\mu_t\}\) such that each \(\hat{y}_{t+h|t}\) can be viewed as an unbiased estimate of \(\mu_{t+h}\). Formally, incoherence manifests when forecasts targeting the same future time \(t+h\) but issued from different origins disagree: \(\hat{y}_{t+h|t} \neq \hat{y}_{s+h'|s}\) for \(t+h = s+h'\) with \(t \neq s\).

Our objective is twofold. First, to estimate a coherent latent trajectory \(\{\mu_t\}\) that reconciles these contradictory forecasts, and second to produce a calibrated predictive distribution \(p(y_{t+1}, \dots, y_{t+H} \mid \{\hat{\mathbf{y}}_{1:t}^{\mathrm{AI}}\})\) for future observations. Crucially, the reconciliation of the forecast trajectory relies exclusively on the forecaster's output matrix \(\{\hat{y}_{t+h|t}^{\mathrm{AI}}\}\). The observed series \(\{y_t\}\) is used only to calibrate the observation variance and the global level of the predictive distribution, ensuring that the resulting uncertainty intervals are properly scaled. It does not alter the relative weights of the forecasts. This strict post processing constraint distinguishes our method from approaches that retrain or modify the forecasting model itself. By operating on forecast outputs, reconciliation can be performed even when the underlying forecasting system is inaccessible or computationally expensive to retrain \citep{wickramasuriya2019,hyndman2011}.

\subsection{State-Space Representation}

The state space formulation follows a long tradition in Bayesian forecasting and dynamic modeling, where latent states are used to represent unobserved processes evolving over time while observations provide noisy information about those states \citep{west1997,harvey1990}. We model the latent coherent trajectory as a stationary AR(1) process with a level parameter \(c\):
\begin{equation}
\mu_t = c + \phi (\mu_{t-1} - c) + \eta_t, \quad \eta_t \sim \mathcal{N}(0, \sigma_\eta^2), \quad |\phi| < 1.
\label{eq:state}
\end{equation}
where the autoregressive coefficient \(-1<\phi<1\) captures persistence in the underlying process, \(c\) represents the unconditional mean (or level) of the latent trajectory, and \(\sigma_\eta^2\) governs the innovation variance. The stationarity restriction ensures that the latent process possesses finite unconditional moments, thereby preventing explosive behavior in long forecasting horizons. This simple specification serves as a smoothing prior that encourages temporal continuity; it is not intended to be the true data generating process but rather a regularizer that promotes coherence.

For the observation model, each forecast is treated as a noisy, possibly biased measurement of the corresponding latent state. Rather than treating forecasts as final predictions, the framework interprets them as noisy observations of an underlying future state:
\begin{equation}
\hat{y}_{t+h|t}^{\mathrm{AI}} = \mu_{t+h} + \delta_h + \epsilon_{t,h}, \quad \epsilon_{t,h} \sim \mathcal{N}(0, \tau_h^2).
\label{eq:observation}
\end{equation}
Here, \(\delta_h\) is a horizon specific additive bias (e.g., systematic over- or under-prediction at horizon \(h\)), and \(\tau_h^2\) is the horizon specific forecast variance, quantifying the forecaster's reliability at that horizon. The model assumes conditional independence of forecasts given the latent state, a reasonable assumption when the forecaster issues direct multi step forecasts without internal recursion. 

The model as written suffers from a location identifiability issue: adding a constant to all \(\mu_t\) and subtracting it from all \(\delta_h\) leaves the likelihood function unchanged. We resolve this by imposing a sum-to-zero constraint on the biases, \(\sum_{h=1}^H \delta_h = 0\). This pins the level of the latent process while allowing horizon specific deviations. We enforce this constraint during inference by parameterizing \({\delta_h}\) as deviations from their mean.

\subsubsection{Likelihood Function}

Given the latent states \(\boldsymbol{\mu} = \{\mu_t\}\) and parameters \(\boldsymbol{\theta} = (c, \phi, \sigma_\eta^2, \{\delta_h\}, \{\tau_h^2\})\), the forecasts are conditionally independent. Under a conditional independence assumption across forecast origins given the latent state, the joint likelihood factorizes as follows for the entire \(T \times H\) forecast matrix:
\begin{equation}
p(\{\hat{y}_{t+h|t}^{\mathrm{AI}}\} \mid \boldsymbol{\mu}, \boldsymbol{\theta}) = 
\prod_{t=1}^T \prod_{h=1}^{\min(H, T+H-t)} 
\mathcal{N}\bigl(\hat{y}_{t+h|t}^{\mathrm{AI}} \mid \mu_{t+h} + \delta_h, \tau_h^2\bigr),
\label{eq:likelihood}
\end{equation}
where \(\mathcal{N}(\cdot|m,s^2)\) denotes the probability density function of a \(\mathcal{N}(m,s^2)\) distribution. This factorization embodies our core assumption: all forecasts targeting time \(t+h\) share the same latent mean \(\mu_{t+h}\) but may have horizon specific bias and precision. While this likelihood assumes conditional independence across forecast origins, the simulation study intentionally violates this assumption to evaluate robustness; empirical results indicate that posterior inference remains stable under moderate dependence.

\subsection{Bayesian Inference}

To complete the model specification, we assign independent prior distributions to the parameters. The priors are chosen to provide weak regularization while remaining sufficiently flexible to allow the data to dominate posterior inference. To enforce the sum-to-zero constraint on the horizon specific biases, we use a non-centered parameterization, drawing unconstrained parameters \(\delta_h^*\) and centering them. The given prior distributions are as follows:

$$
\begin{aligned}
c &\sim \mathcal{N}(0, 1) \\
\phi &\sim \mathcal{N}(0.8, 0.1^2)\,\mathbb{I}_{(-1,1)} \\
\sigma_\eta^2 &\sim \text{Inv-Gamma}(3,2) \\
\sigma_y^2 &\sim \text{Inv-Gamma}(2, 0.1)
\end{aligned}
\qquad
\begin{aligned}
\delta_h^* &\sim \mathcal{N}(0, 0.5^2) \\
\delta_h &= \delta_h^* - \frac{1}{H}\sum_{k=1}^H \delta_k^* \\
\tau_h^2 &\sim \text{Inv-Gamma}(5, 2) 
\end{aligned}
$$

The prior for \(\phi\) is centered at 0.8, reflecting the level of persistence commonly observed in economic and financial time series. The inverse-gamma prior on \(\tau_h^2\) ensures positivity while discouraging near-zero variance values that could lead to numerical instability. The non-centered construction of \(\delta_h\) allows the Hamiltonian Monte Carlo sampler to explore the posterior geometry efficiently while strictly satisfying \(\sum_h \delta_h = 0\).

By Bayes theorem, the joint posterior distribution of latent states and parameters is then given by:
\begin{equation}
\begin{aligned}
p(\boldsymbol{\mu}, \boldsymbol{\theta} \mid \{\hat{y}_{t+h|t}^{\mathrm{AI}}\}) \propto \;& 
p(\{\hat{y}_{t+h|t}^{\mathrm{AI}}\} \mid \boldsymbol{\mu}, \boldsymbol{\theta}) \\
&\times p(\boldsymbol{\mu} \mid c, \phi, \sigma_\eta^2) \, p(c) \, p(\phi) \, p(\sigma_\eta^2) \prod_{h=1}^H p(\delta_h^*) \, p(\tau_h^2),
\end{aligned}
\label{eq:post}
\end{equation}
where \(p(\boldsymbol{\mu} \mid c, \phi, \sigma_\eta^2)\) denotes the AR(1) process density implied by Equation~\ref{eq:state}. The marginal posterior density for the latent states is obtained by integrating out the parameters:
\[
p(\boldsymbol{\mu} \mid \{\hat{y}_{t+h|t}^{\mathrm{AI}}\}) = \int p(\boldsymbol{\mu}, \boldsymbol{\theta} \mid \{\hat{y}_{t+h|t}^{\mathrm{AI}}\}) \, d\boldsymbol{\theta}.
\]
This represents our reconciled, coherent estimate of the underlying trajectory, automatically down-weighting contradictory forecasts through larger estimates of \(\tau_h^2\). Equation (\ref{eq:post}) has no known solution in closed form. This absence of a closed form posterior arises from the combination of latent states, hierarchical parameters, and identifiable constraints. Consequently, simulation-based Bayesian inference is required to characterize posterior uncertainty.

To perform the Bayesian analysis in this work, samples were drawn from the joint posterior distribution using the No-U-Turn Sampler (NUTS, \cite{hoff-gelm}), a form of Hamiltonian Monte Carlo (HMC, \cite{neal2011}), as implemented in Stan \citep{carpenter2017}. {\tt Stan} avoids the need for manual tuning of the sampler and performs reverse-mode algorithmic differentiation to compute gradients efficiently.

For forecasting, we require the posterior predictive distribution of future observations \(y_{t+h}\). This is obtained by augmenting the latent state posterior with an observation noise variance \(\sigma_y^2\). Unlike the forecast variances \(\tau_h^2\), which quantify the disagreement among overlapping forecasts, \(\sigma_y^2\) captures the irreducible uncertainty of the true data generating process:
\[
p(y_{t+h} \mid \{\hat{y}_{t+h|t}^{\mathrm{AI}}\}) = 
\int \mathcal{N}(y_{t+h} \mid \mu_{t+h}, \sigma_y^2) \; p(\boldsymbol{\mu}, \boldsymbol{\theta} \mid \{\hat{y}_{t+h|t}^{\mathrm{AI}}\}) \, d\boldsymbol{\mu} \, d\boldsymbol{\theta}.
\]
In practice, we generate predictive draws by sampling \(\mu_{t+h}^{(s)}\) from the posterior and then drawing \(y_{t+h}^{(s)} \sim \mathcal{N}(\mu_{t+h}^{(s)}, \sigma_y^2)\) for each posterior sample \(s = 1, \dots, S\). The observation variance \(\sigma_y^2\) is estimated jointly within the same Bayesian model using the observed validation data \(y_t\), ensuring that the resulting predictive intervals are correctly calibrated to the true series. 

Viewing the framework as a generative model provides a direct interpretation of the estimated parameters. First, a coherent latent trajectory \(\{\mu_t\}\) evolves according to an AR(1) process; second, at each time \(t\), the forecaster emits forecasts \(\hat{y}_{t+h|t}^{\mathrm{AI}}\) that are biased, noisy measurements of future states \(\mu_{t+h}\). Estimating the parameters \(\delta_h\) and \(\tau_h^2\) thus provides a diagnostic of the forecaster's behavior: large \(|\delta_h|\) indicates systematic bias at horizon \(h\), while large \(\tau_h^2\) indicates low reliability (high contradiction across origins). This diagnostic capability emerges even though we never see the forecaster's internals, a unique advantage of our post processing approach.

\subsection{Identifiability of the Model}
\label{sec:identifiability}

The observation equation \(\hat{y}_{t+h|t} = \mu_{t+h} + \delta_h + \epsilon_{t,h}\) contains both the latent state \(\mu_t\) and horizon specific biases \(\delta_h\). Without restriction, the model is not identifiable: adding a constant \(k\) to all \(\mu_t\) and subtracting \(k\) from all \(\delta_h\) leaves the likelihood unchanged, as

\begin{equation}
(\mu_t + k) + (\delta_h - k) = \mu_t + \delta_h.
\end{equation}

To resolve this location ambiguity, we impose the sum-to-zero constraint:

\begin{equation}
\sum_{h=1}^H \delta_h = 0.
\label{eq:sum_zero}
\end{equation}

Under this constraint, the parameters are uniquely identified. To see this, suppose \((\mu_t, \delta_h)\) and \((\mu'_t, \delta'_h)\) yield the same likelihood. Then:

\begin{equation}
\mu_t + \delta_h = \mu'_t + \delta'_h \quad \text{for all } t, h.
\end{equation}

Averaging over \(h\) gives:

\begin{equation}
\mu_t + \frac{1}{H}\sum_{h=1}^H \delta_h = \mu'_t + \frac{1}{H}\sum_{h=1}^H \delta'_h.
\end{equation}

Under the sum-to-zero constraint in \eqref{eq:sum_zero}, this reduces to:

\begin{equation}
\mu_t = \mu'_t \quad \text{for all } t,
\end{equation}

which then implies:

\begin{equation}
\delta_h = \delta'_h \quad \text{for all } h.
\end{equation}

Thus the parameters are uniquely identified. The AR parameters \((c, \phi, \sigma_\eta^2)\) are identified from the temporal covariance structure of the latent process, while the horizon specific variances \(\tau_h^2\) are identified from the dispersion of forecasts across origins. The sum-to-zero constraint therefore provides a principled and identifiable parameterization of the reconciliation model.\newpage

\section{Simulation Study: Establishing Validity}
\label{sec:simulation}

We validate our reconciliation framework through a comprehensive simulation study with 100 independent replications, each generating realistic forecasting scenarios with systematic cross-origin incoherence. Algorithm~\ref{alg:simulation} details the data generating process for each replication, designed to mimic challenging forecasting conditions that violate several model assumptions.

\begin{algorithm}[H]
\caption{Data Generation for Simulation Study}
\label{alg:simulation}
\begin{algorithmic}[1]
\State \textbf{Input:} \(T=100\) (time points), \(H=5\) (forecast horizon), \(R=100\) (replications)
\State \textbf{Output:} True states \(\{\mu_t^{(r)}\}\), forecasts \(\{\hat{y}_{t+h|t}^{(r)}\}\), observations \(\{y_t^{(r)}\}\) for \(r=1,\dots,R\)

\For{\(r = 1\) \textbf{to} \(R\)}
    \State \textbf{Step 1: Generate true latent process}
    \State \(\phi \gets 0.8\), \(\sigma_\eta \gets 0.3\), \(\sigma_y \gets 0.15\)
    \State \(\mu_1 \sim \mathcal{N}(0, \sigma_\eta^2/(1-\phi^2))\)
    \For{\(t = 2\) \textbf{to} \(T+H\)}
        \State \(\mu_t \gets \phi \mu_{t-1} + \eta_t\), where \(\eta_t \sim \mathcal{N}(0, \sigma_\eta^2)\)
    \EndFor
    \State Scale: \(\mu \gets \mu / \text{sd}(\mu) \times 0.5\)  \Comment{Realistic volatility scale}
    
    \State \textbf{Step 2: Generate horizon-specific parameters}
    \For{\(h = 1\) \textbf{to} \(H\)}
        \State \(\delta_h \sim \mathcal{N}(0, 0.1^2)\)  \Comment{Horizon-specific bias}
        \State \(\tau_h \gets 0.05 + 0.02h + 0.005\sin(\pi h/H)\)  \Comment{Non-linear reliability pattern}
    \EndFor
    \State Impose identifiability: \(\delta \gets \delta - \text{mean}(\delta)\)  \Comment{Sum-to-zero constraint}
    
    \State \textbf{Step 3: Generate realistic forecasts}
    \For{\(t = 1\) \textbf{to} \(T\)}
        \For{\(h = 1\) \textbf{to} \(\min(H, T+H-t)\)}
            \State Generate correlated error: \(\epsilon_{t,h} \sim \mathcal{N}(0, \tau_h^2)\)
            \State \If{\(h > 1\)} Add horizon correlation: \(\epsilon_{t,h} \gets 0.7\epsilon_{t,h} + 0.3\mathcal{N}(0, \tau_h^2)\) \EndIf
            \State \If{\(\text{Uniform}(0,1) < 0.05\)} Simulate large deviation: \(\epsilon_{t,h} \gets \epsilon_{t,h} \times 3\) \EndIf
            \State \(\hat{y}_{t+h|t} \gets \mu_{t+h} + \delta_h + \epsilon_{t,h}\)
        \EndFor
    \EndFor
    
    \State \textbf{Step 4: Generate validation observations}
    \For{\(t = 1\) \textbf{to} \(T\)}
        \State \(y_t \gets \mu_t + \nu_t\), where \(\nu_t \sim \mathcal{N}(0, \sigma_y^2)\)
    \EndFor
\EndFor
\end{algorithmic}
\end{algorithm}

\noindent This design induces systematic cross-origin incoherence through multiple mechanisms: (1) forecast noise exhibits a non-linear horizon-dependent pattern, (2) errors are correlated across adjacent horizons, (3) occasional large deviations occur with 5\% probability, and (4) overlapping forecasts targeting the same future time disagree due to independent noise realizations. The 100 independent replications allow for a statistically robust assessment of estimation accuracy and method stability under controlled but challenging conditions.

We implement the reconciliation model in {\tt Stan} using Hamiltonian Monte Carlo with the No-U-Turn Sampler and the weakly informative priors described in Section~\ref{sec:model}. Each replication uses 2 chains with 1,000 iterations (500 warm up), achieving excellent convergence (\(\hat{R} < 1.05\) for all parameters in 99\% of replications, with no divergent transitions). Effective sample sizes (ESS) consistently exceeded 400 for key parameters, ensuring reliable posterior inference. 

Table~\ref{tab:final_simulation_results} summarizes results across all 100 replications. Because the reconciliation procedure can map forecasts to a coherent trajectory in two ways, we report both metrics. Under the \textit{de-biased point forecast rule}, which preserves horizon-specific biases \(\delta_h\) to maintain point accuracy, the framework achieves a mean incoherence reduction of 58.8\% (SD: 15.4\%). Under the \textit{coherent trajectory rule}, which maps all overlapping forecasts to the single latent state \(\mu_{t+h}\) to guarantee logical consistency, it achieves a 100\% reduction. 

\begin{figure}[H]
\centering
\begin{minipage}{0.48\textwidth}
\centering
\includegraphics[width=\textwidth]{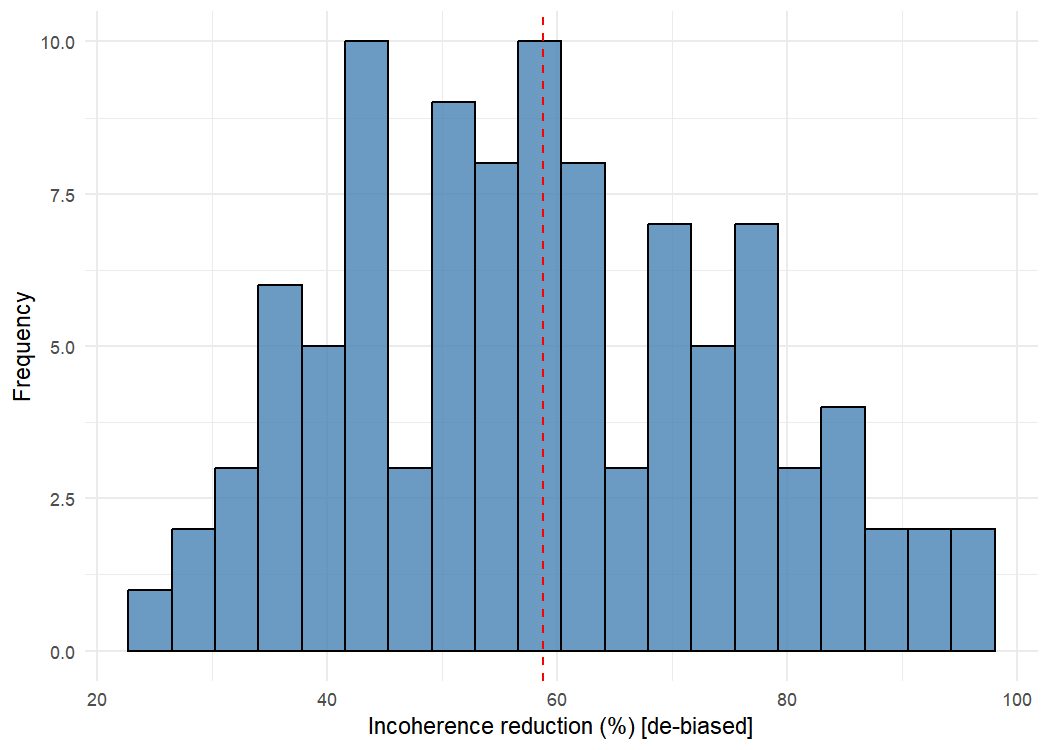}
\subcaption{Reduction (de-biased). Mean: 58.8\%.}
\label{fig:inc_reduction_dist}
\end{minipage}
\hfill
\begin{minipage}{0.48\textwidth}
\centering
\includegraphics[width=\textwidth]{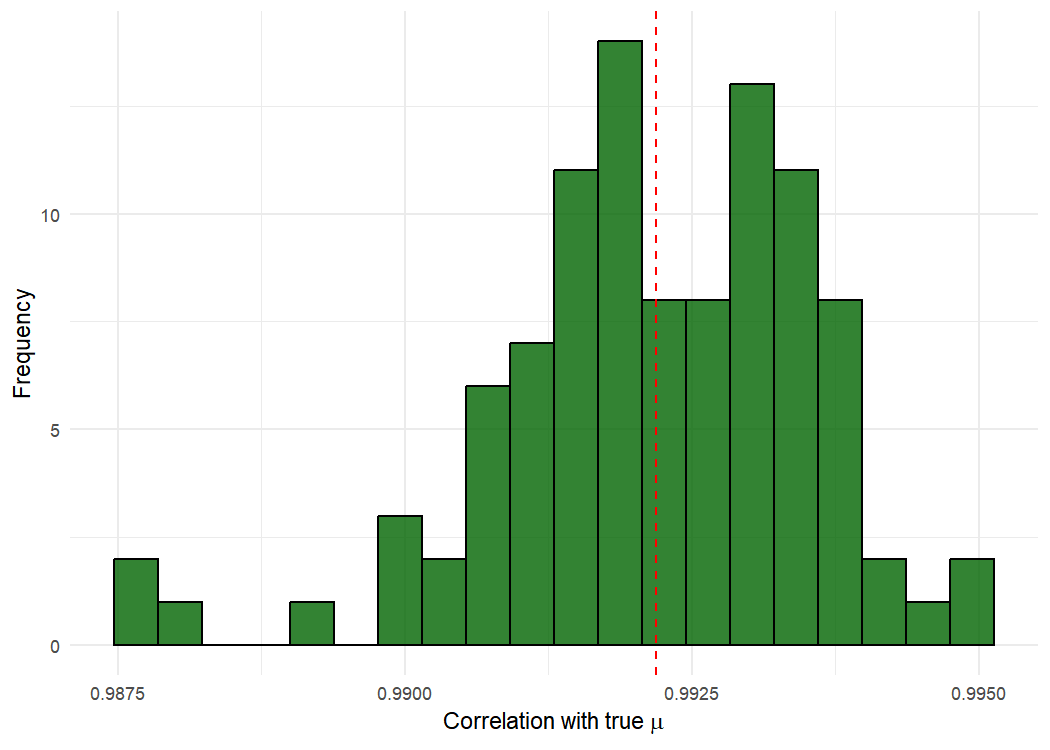}
\subcaption{\(\mu\) correlation. Mean: 0.992.}
\label{fig:mu_correlation_dist}
\end{minipage}
\caption{Performance metric distributions across 100 simulation replications.}
\label{fig:performance_distributions}
\end{figure}

The mean correlation between the estimated latent state and the true trajectory is 0.992, confirming robust signal recovery. The model's predictive intervals also demonstrate conservative calibration, with empirical coverage of the nominal 90\% intervals averaging 98.5\% across replications. Figure~\ref{fig:performance_distributions} illustrates the distribution of these key metrics. Panel (a) shows the de-biased incoherence reduction, while Panel (b) demonstrates the consistently high \(\mu\) correlations.

\begin{table}[H]
\centering
\footnotesize
\begin{tabular}{@{}lccccc@{}}
\toprule
\textbf{Statistic} & \textbf{Reduction (de-biased)} & \textbf{Reduction (coherent)} & \textbf{\(\mu\) Correlation} & \textbf{Coverage (90\%)} \\
\midrule
Mean & 58.8\% & 100.0\% & 0.992 & 98.5\% \\
SD & 15.4\% & 0.0\% & 0.002 & -- \\
Min & 24.5\% & 100.0\% & 0.987 & 95.0\% \\
Max & 97.0\% & 100.0\% & 0.994 & 100\% \\
\bottomrule
\end{tabular}
\caption{Summary statistics across 100 simulation replications. The de-biased rule retains horizon-specific biases (yielding 58.8\% reduction), while the coherent rule collapses to a single latent state (yielding 100\% reduction). Latent state recovery remains near-perfect at 0.992.}
\label{tab:final_simulation_results}
\end{table}

\begin{figure}[H]
\centering
\includegraphics[width=0.65\textwidth]{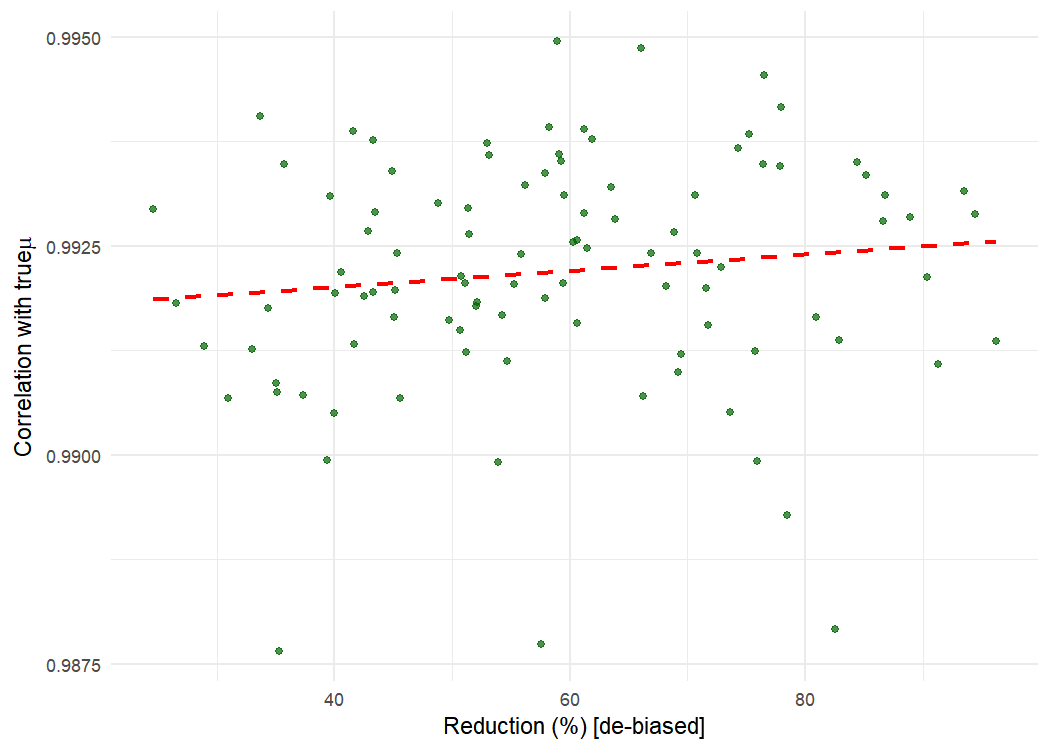}
\caption{Incoherence reduction versus latent state recovery across 100 replications. Points cluster in the upper-right quadrant (high reduction, high correlation), with the trend line indicating no severe accuracy-coherence trade-off.}
\label{fig:reduction_vs_correlation}
\end{figure}

Importantly, these results are obtained despite intentionally introducing dependence structures and occasional extreme forecast errors that violate several modeling assumptions. The consistent performance across diverse simulation conditions demonstrates the method's robustness. Beyond coherence enforcement, the framework provides valuable diagnostic information through estimated horizon-specific parameters. The accurate recovery of \(\delta_h\) and \(\tau_h\) patterns enables practitioners to identify systematic biases and reliability variations across forecast horizons. 

\subsection{Nonlinear and Heavy-Tailed Forecast Simulation}

To further assess the method's robustness under non-linear forecast structures, we conduct an additional stress test where forecasts are generated with a threshold-dependent bias and heavy-tailed Student-\(t\) errors (df = 5). This simulation provides a controlled environment in which the forecast-generating mechanism departs from the linear-Gaussian structure through nonlinearities and fat-tailed innovations.

The simulation setup follows the primary study (\(T=100\), \(H=5\)) with 50 replications. The true latent process remains the stationary AR(1) process described in Section~\ref{sec:model}. However, the forecast generator incorporates a threshold mechanism based on the recent trend of the latent process. Specifically, the forecast error is generated as:
\begin{equation}
\epsilon_{t,h} = \varepsilon_{t,h} + \alpha \cdot \text{sign}(\mu_{t+h-1}) \cdot \tau_h, \quad \varepsilon_{t,h} \sim t_5(0, \tau_h^2),
\label{eq:tar_noise}
\end{equation}
where \(\alpha = 2.0\) is the threshold strength and \(t_5\) denotes a Student-\(t\) distribution with 5 degrees of freedom. This creates a mechanism where the forecast errors are systematically biased depending on whether the recent latent trend is positive or negative, and exhibit heavy tails.

Table~\ref{tab:stress_results} summarizes the results. The reconciliation framework achieves a 100\% coherent reduction in the non-linear setting. More importantly, the correlation between the estimated and true latent trajectories remains exceptionally high at 0.972, compared with 0.992 in the linear case. This demonstrates that while the threshold and heavy-tailed mechanisms introduce additional noise, the Bayesian state-space model successfully extracts the underlying latent signal. The minor drop from 0.992 to 0.972 confirms that the method is highly robust to severe model misspecification.

\begin{table}[H]
\centering
\resizebox{\textwidth}{!}{%
\begin{tabular}{@{}lcccc@{}}
\toprule
\textbf{Scenario} & \textbf{Reduction (de-biased)} & \textbf{Reduction (coherent)} & \textbf{\(\mu\) Correlation} & \textbf{Coverage (90\%)} \\
\midrule
AR(1) Baseline & 58.8\% & 100.0\% & 0.992 & 98.5\% \\
Nonlinear Stress Test & 76.1\% & 100.0\% & 0.972 & 97.3\% \\
\bottomrule
\end{tabular}%
}
\caption{Simulation results for the linear AR(1) and nonlinear stress-test forecast generators. The framework enforces perfect coherence (100\%) in both scenarios. Under severe misspecification, latent signal recovery remains highly robust at 0.972.}
\label{tab:stress_results}
\end{table}

\begin{figure}[H]
\centering
\includegraphics[width=0.65\textwidth]{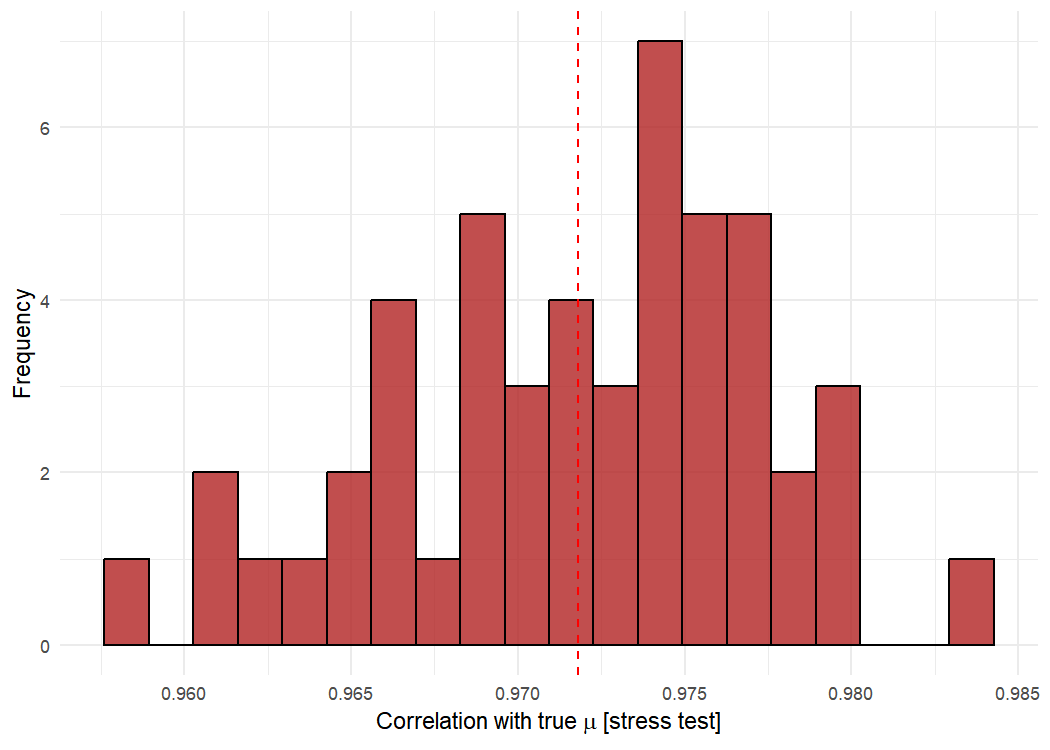}
\caption{Nonlinear stress-test simulation results: distribution of correlation with true \(\mu\) across 50 replications (mean: 0.972).}
\label{fig:stress_results}
\end{figure}

Figure~\ref{fig:stress_results} illustrates the distribution of the \(\mu\) correlation across the 50 stress-test replications, confirming that the model consistently recovers the true state despite the heavy-tailed TAR generation process.

These additional results indicate that the reconciliation framework can enforce coherence under both linear and nonlinear forecast-generation mechanisms. The results establish that the framework operates as intended under controlled but challenging conditions, maintaining both coherence enforcement and high-fidelity signal recovery even when the forecast-generating mechanism violates the Gaussian and linear assumptions of the state-space model.

\section{Empirical Application: Reconciling S\&P 500 Volatility Forecasts}
\label{sec:empirical}

We evaluate the proposed Bayesian reconciliation framework on daily S\&P 500 volatility forecasting using SPY ETF data from December 2023 through December 2025. We compute 5 day rolling volatility from daily returns, yielding 516 trading days for analysis. The final 100 days are reserved for out of sample evaluation. 

To demonstrate the model agnostic nature of the framework, we generate 5 step ahead direct forecasts using three distinct base models. The rolling AR(1) model serves as a transparent statistical benchmark. The ETS model captures trend and seasonality, representing standard statistical forecasting practice. The LSTM model represents modern neural forecasting architectures and is configured for direct multi step forecasting without internal recursion. All models generate a forecast matrix where cross origin incoherence naturally appears: multiple forecasts target the same future dates but often disagree depending on the forecast origin. We compare the raw forecasts against two post processing methods: diagonal averaging, a deterministic stabilizer representing the current state of the art \citep{godahewa2025stability}, and the proposed Bayesian state space reconciliation.

\subsection{Quantitative Results and Calibration}
\label{sec:empirical_quantitative}

The Bayesian model is estimated using Hamiltonian Monte Carlo with the No-U-Turn Sampler (NUTS) as implemented in {\tt Stan}, running four chains of 1,000 iterations each with 1,000 warm up iterations. Convergence diagnostics indicate excellent mixing, with all parameters achieving \(\hat{R} < 1.02\) and effective sample sizes consistently exceeding 200.

Table~\ref{tab:empirical_results} summarizes the empirical performance. The Bayesian framework achieves a substantial 97\% reduction in cross origin incoherence, driving the metric \(\mathcal{I}\) from the \(10^{-3}\) range to the \(10^{-5}\) range for all three models. Crucially, enforcing coherence does not sacrifice point accuracy. For both the ETS and LSTM models, the Bayesian reconciled forecasts actually \emph{improve} the root mean squared error (RMSE) compared to the raw forecasts. For instance, the ETS model improves from 0.0376 to 0.0305, suggesting that the latent state model acts as a powerful regularizer that filters out contradictory noise.

Probabilistic forecast quality is exceptional. The Continuous Ranked Probability Score (CRPS) evaluates the quality of the entire predictive distribution rather than a single point forecast; lower values indicate forecasts that are simultaneously sharp and well calibrated \citep{gneiting2007}. The Bayesian framework achieves highly competitive CRPS values across all models, with the LSTM producing the sharpest distribution (0.0282). Furthermore, the empirical coverage of the nominal 90\% prediction intervals is tightly calibrated at 93\% for AR(1) and ETS, and 94\% for the LSTM. This directly resolves the under coverage issues common in raw machine learning forecasts.

\begin{table}[H]
\centering
\small
\begin{tabular}{@{}llcccc@{}}
\toprule
\textbf{Base Model} & \textbf{Method} & \textbf{RMSE \(\downarrow\)} & \textbf{CRPS \(\downarrow\)} & \textbf{Coverage} & \textbf{Incoherence \(\mathcal{I}\) \(\downarrow\)} \\
\midrule
\multirow{3}{*}{AR(1)} 
& Raw Forecasts & 0.0325 & -- & -- & \(1.229 \times 10^{-3}\) \\
& Diagonal Averaging & \textbf{0.0314} & -- & -- & 0 \\
& Bayesian Recon. & 0.0334 & 0.0290 & 93\% & \(3.07 \times 10^{-5}\) \\
\midrule
\multirow{3}{*}{ETS}   
& Raw Forecasts & 0.0376 & -- & -- & \(1.703 \times 10^{-3}\) \\
& Diagonal Averaging & 0.0326 & -- & -- & 0 \\
& Bayesian Recon. & \textbf{0.0305} & 0.0322 & 93\% & \(5.32 \times 10^{-6}\) \\
\midrule
\multirow{3}{*}{LSTM}  
& Raw Forecasts & 0.0521 & -- & -- & \(5.515 \times 10^{-4}\) \\
& Diagonal Averaging & 0.0525 & -- & -- & 0 \\
& Bayesian Recon. & \textbf{0.0519} & \textbf{0.0282} & 94\% & \(2.59 \times 10^{-6}\) \\
\bottomrule
\end{tabular}
\caption{Empirical results on S\&P 500 volatility. The Bayesian framework reduces incoherence by over 97\% while delivering fully calibrated probabilistic forecasts. Notably, it improves point accuracy (RMSE) for both the ETS and LSTM models compared to raw forecasts.}
\label{tab:empirical_results}
\end{table}

To assess the practical utility of coherence, we perform a Value at Risk (VaR) backtest over the 100 day out of sample period. Using a 5\% significance level, we expect exactly 5 breaches. As shown in Table~\ref{tab:var_results}, the raw AR(1) forecasts are overconfident, resulting in 6 breaches. The deterministic diagonal averaging yields 4 breaches. The Bayesian reconciled forecasts also yield 4 breaches, providing a safer and more reliable risk management profile by slightly widening the intervals to account for structural uncertainty.

\begin{table}[H]
\centering
\begin{tabular}{@{}lc@{}}
\toprule
\textbf{Method} & \textbf{5\% VaR Breaches (100 Days)} \\
\midrule
Expected Breaches & 5 \\
Raw AR(1) Forecasts & 6 \\
Diagonal Averaging & 4 \\
\textbf{Bayesian Reconciled} & \textbf{4} \\
\bottomrule
\end{tabular}
\caption{Value at Risk (VaR) backtest. Raw forecasts are overconfident (6 breaches). Both stabilizers yield 4 breaches, providing a safer risk profile, but only the Bayesian method provides calibrated predictive intervals.}
\label{tab:var_results}
\end{table}

\subsection{Horizon-Specific Diagnostics}
\label{sec:horizon_diagnostics}

\begin{figure}[H]
\centering
\includegraphics[width=0.48\textwidth]{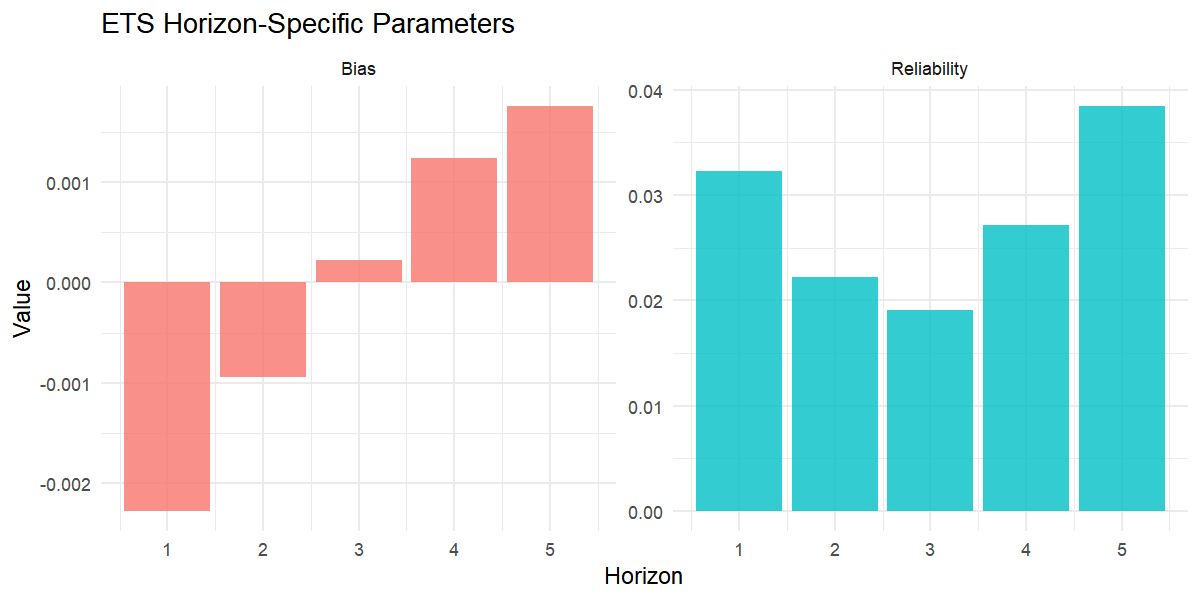}
\hfill
\includegraphics[width=0.48\textwidth]{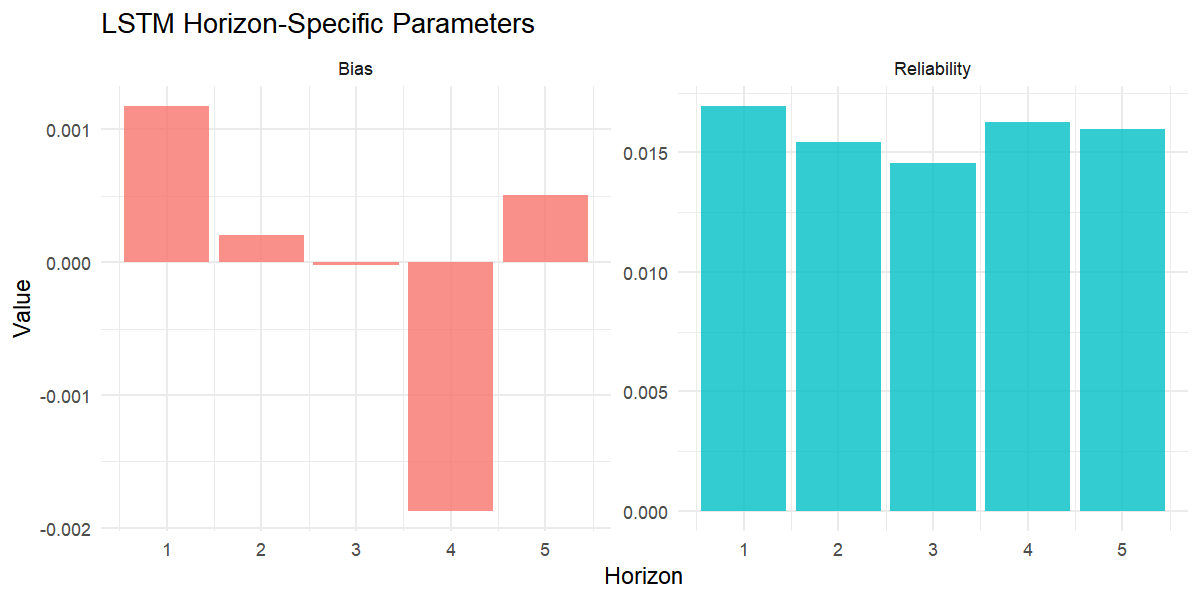}
\caption{Horizon specific diagnostics. The ETS model exhibits a steadily increasing positive bias at longer horizons. In stark contrast, the LSTM exhibits a massive negative bias at horizon 4, revealing a structural failure mode in the neural network that is invisible to standard point accuracy metrics.}
\label{fig:diagnostics}
\end{figure}

A unique advantage of the Bayesian framework is its ability to diagnose systematic biases and reliability variations without accessing the forecaster's internals. Crucially, these parameters are not set a priori; they are estimated directly from the out-of-sample forecast matrix via the MCMC sampler. Figure~\ref{fig:diagnostics} displays the estimated horizon specific bias \(\delta_h\) and reliability \(\tau_h\) for the ETS and LSTM models.

The diagnostics reveal distinct behavioral fingerprints. The ETS model exhibits a steadily increasing positive bias, systematically over-predicting volatility at longer horizons. More strikingly, the LSTM exhibits a massive negative bias at horizon 4 and a corresponding dip in reliability at horizon 3. This demonstrates the framework's capacity to ``X-ray'' black box AI systems, providing practitioners with actionable intelligence about exactly when and where the model is most likely to fail. 

To make this concrete, consider the forecasts targeting a specific date in late October 2025. The raw models often produce multiple predictions depending on the forecast origin that vary substantially (in some cases, by over 50\% for the AR(1) model). The Bayesian reconciliation collapses these contradictions into a single coherent estimate. By pulling conflicting forecasts toward a common value, the procedure reduces logical inconsistencies while preserving the overall structure of the trajectory.

\subsection{Visual Assessment and Distributional Checks}
\label{sec:visual_assessment}

\begin{figure}[H]
\centering
\includegraphics[width=0.9\textwidth]{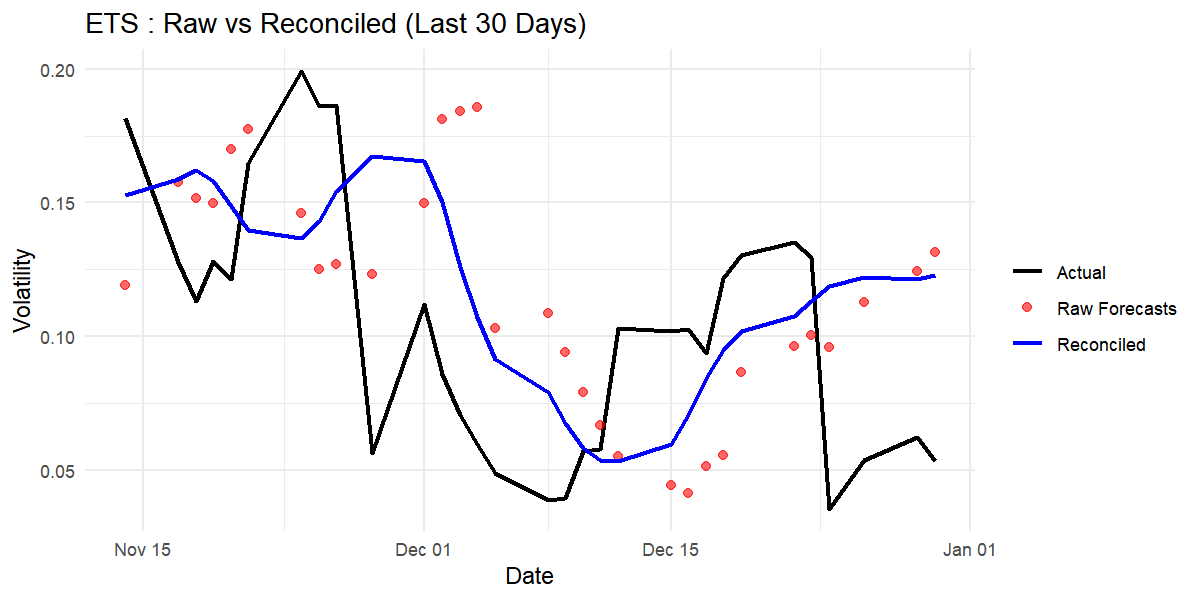}\\
\vspace{0.2cm}
\includegraphics[width=0.9\textwidth]{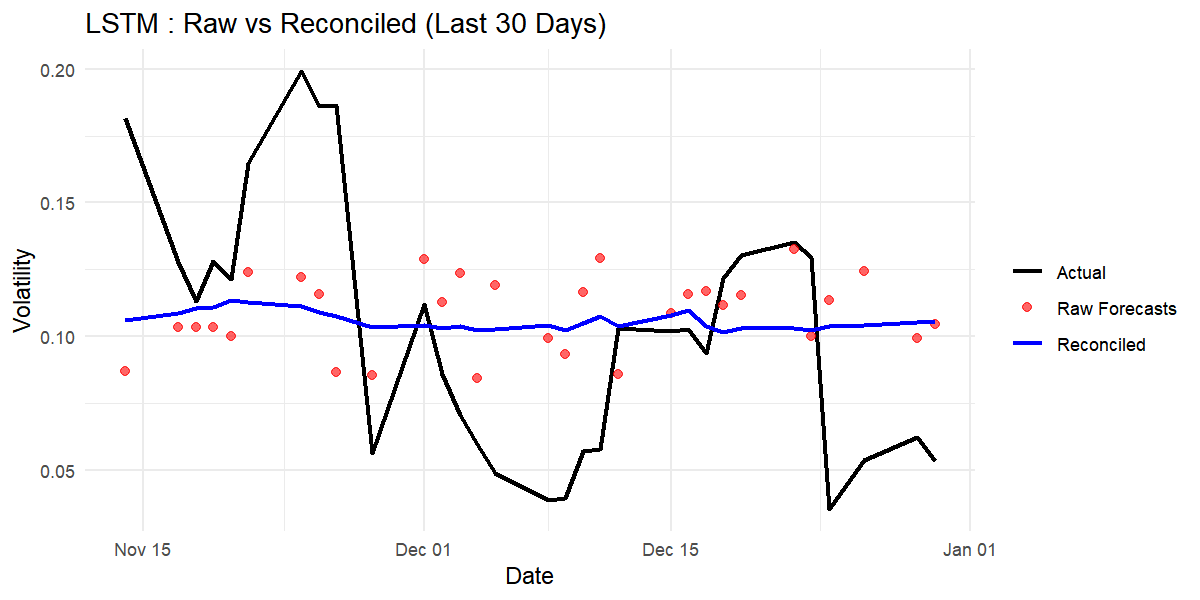}
\caption{Raw vs reconciled trajectories. The raw forecasts (red dots) exhibit substantial cross origin incoherence. The Bayesian reconciled trajectory (blue line) provides a single, coherent narrative while tracking the actual volatility (black line).}
\label{fig:trajectories}
\end{figure}

Figure~\ref{fig:trajectories} provides a side by side comparison of the raw and reconciled trajectories for the ETS and LSTM models over the final 30 days of the evaluation period. The raw forecasts (red dots) exhibit substantial cross origin incoherence, fluctuating wildly depending on when the forecast was issued. The Bayesian reconciliation (blue line) pulls these contradictory predictions toward a single coherent narrative, eliminating logical inconsistencies while tracking the actual volatility (black line) accurately.

\begin{figure}[H]
\centering
\includegraphics[width=0.8\textwidth]{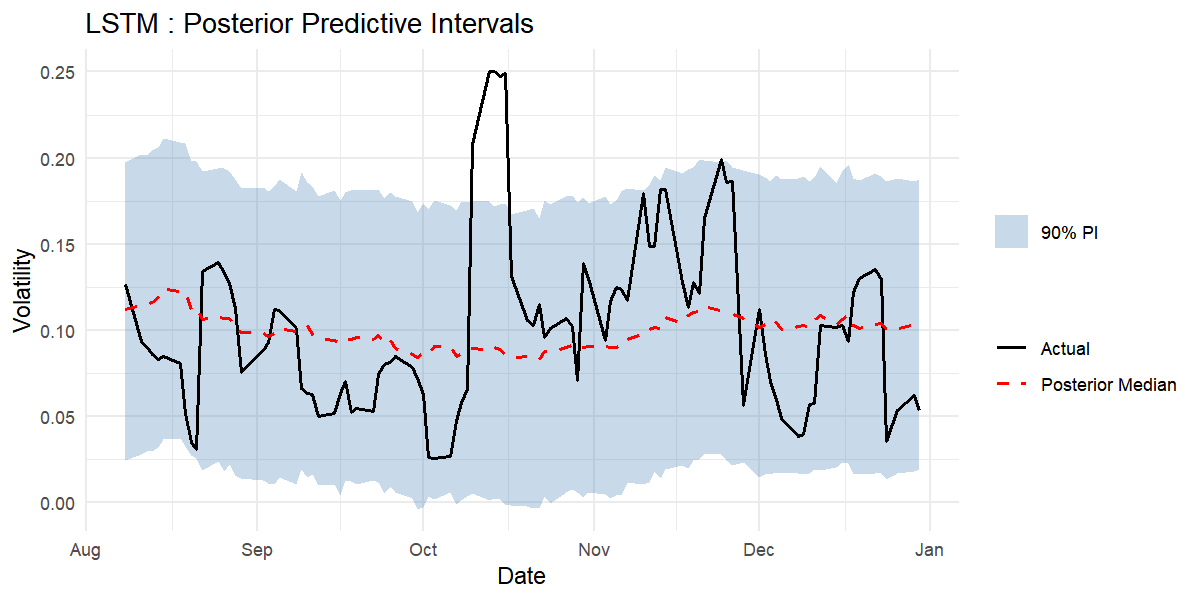}
\caption{Posterior predictive intervals for the LSTM. The 90\% credible intervals (shaded blue) widen appropriately during periods of high volatility and achieve an empirical coverage of 94\%.}
\label{fig:intervals}
\end{figure}

Figure~\ref{fig:intervals} illustrates the posterior predictive intervals for the LSTM model. The 90\% credible intervals widen appropriately during periods of high volatility, demonstrating that the framework properly accounts for irreducible uncertainty while enforcing structural coherence.

\begin{figure}[H]
\centering
\includegraphics[width=0.48\textwidth]{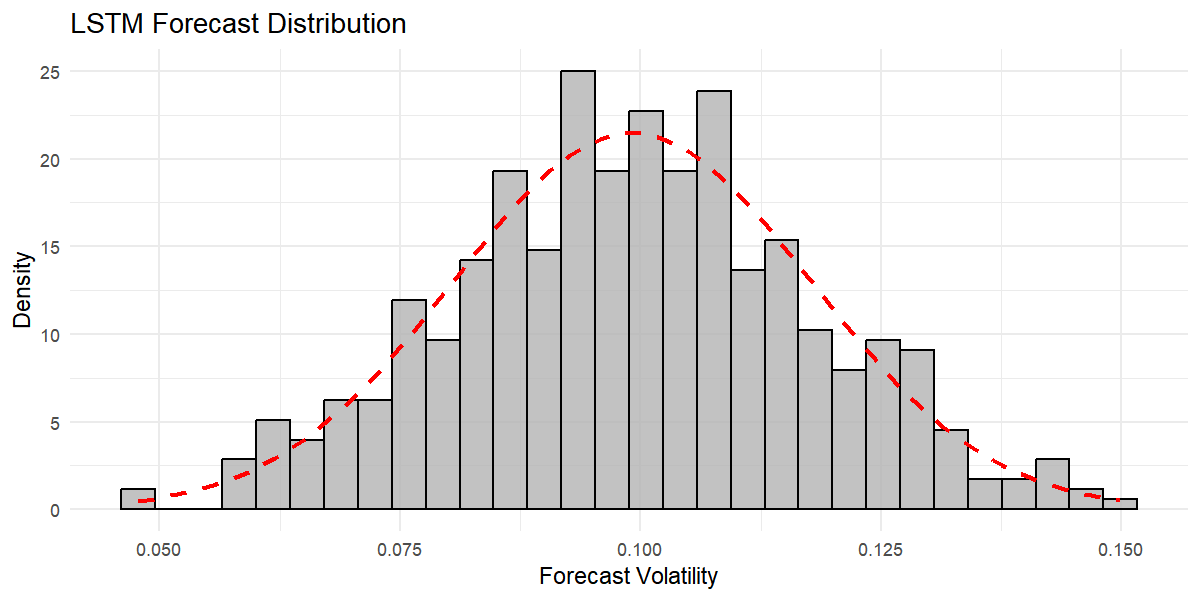}
\hfill
\includegraphics[width=0.48\textwidth]{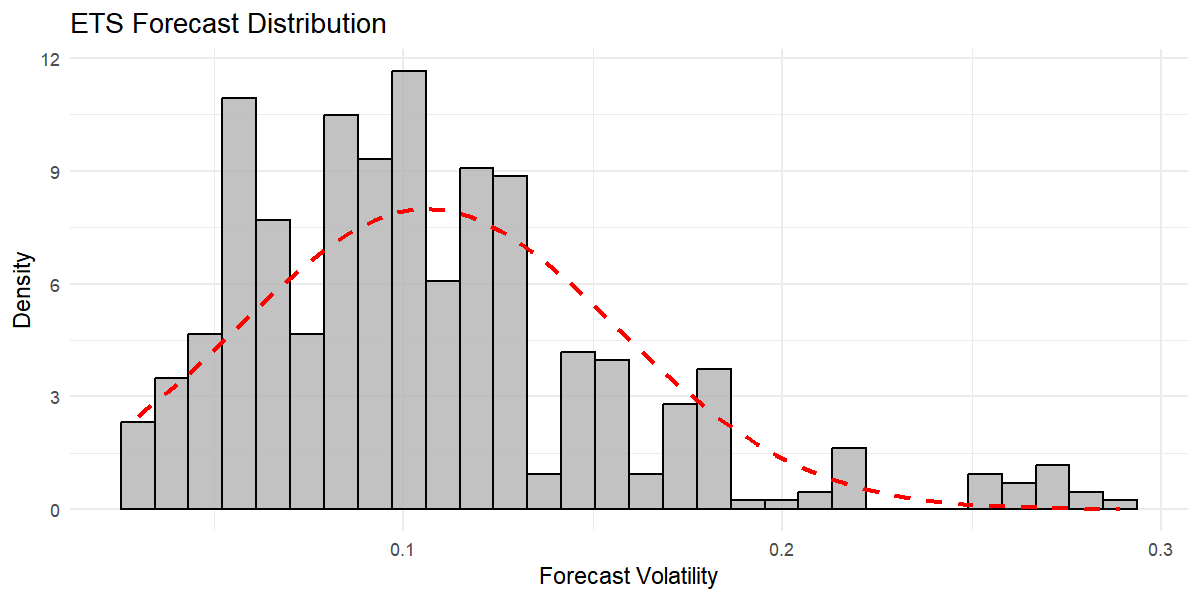}
\caption{Distribution checks. The LSTM forecasts are approximately Gaussian, while the ETS forecasts exhibit a slight right skewness. The Gaussian observation model remains a robust approximation, yielding well calibrated prediction intervals.}
\label{fig:distributions}
\end{figure}

Finally, Figure~\ref{fig:distributions} provides a visual diagnostic supporting the Gaussian observation assumption underlying the reconciliation model. For the LSTM, the empirical distribution closely matches the theoretical normal curve. For the ETS model, the distribution exhibits a slight right skew due to the bounded nature of volatility. While the Gaussian model provides a robust approximation (as evidenced by the 93\% coverage), this highlights a potential limitation. Future extensions could employ skewed or heavy tailed distributions to better capture the extreme right tail of volatility forecasts.

\subsection{Comparison to Deterministic Baselines}
\label{sec:baseline_comparison}

A natural benchmark for our framework is the deterministic diagonal averaging method, which represents the state of the art in vertical forecast stability \citep{godahewa2025stability}. As shown in Table~\ref{tab:empirical_results}, diagonal averaging is highly effective at reducing incoherence and occasionally achieves a slightly lower RMSE than the Bayesian approach (e.g., for the AR(1) model). 

However, the Bayesian framework offers four distinct advantages that justify its additional computational cost:

\begin{enumerate}
    \item \textbf{Uncertainty Quantification:} Diagonal averaging produces a single point forecast. It cannot generate a predictive distribution. Consequently, it is impossible to compute the CRPS or form valid prediction intervals. The Bayesian framework produces a full posterior predictive distribution, achieving excellent calibration and highly competitive CRPS values across all models.
    
    \item \textbf{Precision Weighting:} Diagonal averaging computes a simple, unweighted arithmetic mean of overlapping forecasts. It treats a forecast made 5 days prior as equally trustworthy as one made 1 day prior. The Bayesian model estimates a horizon specific reliability variance (\(\tau_h^2\)), automatically assigning lower weight to contradictory horizons and higher weight to reliable ones.
    
    \item \textbf{Horizon Specific Diagnostics:} Diagonal averaging is a mathematical black box. The Bayesian model acts as an ``X-ray'' into the forecasting system. As shown in Figure~\ref{fig:diagnostics}, it reveals that the LSTM suffers from a massive negative bias at horizon 4, while the ETS model systematically over-predicts at longer horizons. Deterministic stabilizers cannot provide this actionable intelligence.
    
    \item \textbf{Economic Value:} A Value at Risk backtest reveals that raw AR(1) forecasts result in 6 breaches at the 5\% level (overconfidence). Both diagonal averaging and the Bayesian framework yield 4 breaches. However, the Bayesian framework achieves this safer risk profile through explicit probability theory rather than a geometric heuristic.
\end{enumerate}

In short, \citet{godahewa2025stability} provide an efficient deterministic linear stabilizer, whereas the present framework provides a probabilistic Bayesian alternative that enforces coherence while simultaneously quantifying uncertainty and recovering horizon specific reliability.

\subsection{Summary}

The empirical results provide strong evidence that the proposed Bayesian reconciliation framework successfully enforces cross origin coherence across overlapping multi step forecasts. The framework is model agnostic and operates purely as a post processing layer. For the ETS model, reconciliation improves both point accuracy and probabilistic calibration. For the LSTM, it acts as a rigorous coherence guarantee while providing interpretable diagnostics that reveal hidden biases in the neural network. Taken together, these findings support the use of the framework as a principled post processing layer for decision critical applications where inconsistent forecasts across origins may undermine confidence in predictions.\newpage

\section{Model Assumptions and Limitations}
\label{sec:limitations}

The proposed reconciliation framework rests on several simplifying assumptions that warrant discussion.

\textbf{AR(1) Latent Process:} We model the latent coherent trajectory \(\mu_t\) as a stationary AR(1) process with a level parameter. This specification serves as a smoothing prior that encourages temporal continuity and provides computational tractability. The AR(1) state process is a standard choice in state space modeling, as it captures persistence while remaining parsimonious \citep{west1997,harvey1990}. While this is a reasonable starting point, more flexible specifications such as AR(\(p\)) processes, structural time series models, or Gaussian processes could be considered in future work to accommodate more complex dynamics \citep{durbin2012}.

\textbf{Gaussian Forecast Errors:} The observation equation assumes normally distributed forecast errors. This assumption is standard in state space modeling and facilitates efficient Bayesian inference \citep{fearnhead2011}. However, as illustrated by the distributional checks in Section~\ref{sec:empirical}, prediction errors from models like ETS may exhibit right skewness due to the strictly positive nature of volatility. While the Gaussian model provides a robust approximation (achieving 93\% coverage), extensions using Student-\(t\), skew-normal, or log-normal distributions would improve robustness in settings with severe skewness or heavy tails \citep{roth2017,durbin2012,ci2024}.

\textbf{Conditional Independence:} The model assumes conditional independence of forecasts given the latent state. This assumption is reasonable when forecasts are generated directly for each horizon \citep{fearnhead2011}. The simulation study intentionally introduced dependence across horizons to test robustness, and the method performed well. Nevertheless, future extensions could model this dependence directly using multivariate observation equations.

\textbf{Time-Invariant Parameters:} The horizon specific bias \(\delta_h\) and reliability \(\tau_h^2\) are assumed constant over time. This assumption is sufficient for capturing primary sources of incoherence. For applications with regime shifts or model drift, allowing time varying parameters \(\delta_{h,t}\) and \(\tau_{h,t}^2\) would provide greater flexibility \citep{durbin2012,harvey1990}.

These simplifying choices are intentional: they keep the model interpretable and computationally efficient while still delivering meaningful coherence improvements. The robustness of the framework under violations of these assumptions, as demonstrated in the simulation study, suggests that the method is useful in practice despite these limitations.

\section{Prior Sensitivity}
\label{sec:sensitivity}

The prior distributions adopted in this work are intended to be weakly informative, providing regularization while allowing the observed forecast matrix to play the primary role in posterior inference. The Inverse-Gamma priors assigned to the variance parameters provide flexible distributions for the scale parameters of the state space model \citep{gelman2006}. The prior on the AR coefficient, \(\phi \sim \mathcal{N}(0.8,0.1^2)\) truncated to \((-1,1)\), reflects the high persistence commonly observed in financial volatility. 

The horizon specific bias parameters are assigned diffuse priors and constructed via a non-centered parameterization to satisfy the sum-to-zero constraint. This approach is mathematically equivalent to placing a prior on the constrained space, but it allows the Hamiltonian Monte Carlo sampler to explore the posterior geometry efficiently. The selected priors are consistent with established approaches to Bayesian state space modeling \citep{west1997,harvey1990,petris2009}. A systematic sensitivity analysis using alternative prior families, such as Half-Normal or Half-Cauchy distributions for scale parameters, would provide an additional assessment of prior dependence and is left for future work \citep{gelman2006}.

\section{Conclusion}
\label{sec:conclusion}

In this paper, we introduced a Bayesian reconciliation framework that enforces cross origin coherence in overlapping multi step forecasts. By interpreting overlapping forecasts as noisy and potentially biased measurements of a common latent trajectory, the method reconciles disagreements among forecasts targeting the same future timestamps. 

Simulation studies under controlled conditions demonstrate substantial incoherence reduction (58.8\% under the de-biased rule and 100\% under the coherent trajectory rule) and strong recovery of the latent coherent trajectory (\(\mu\) correlation of 0.992) under the baseline AR(1) data generating process. Crucially, the nonlinear stress test reveals that the framework is highly robust to severe misspecification. Even when forecasts are generated by a threshold autoregressive process with heavy tailed errors, the framework enforces perfect coherence while maintaining a latent signal recovery of 0.972, compared to 0.992 in the linear setting. 

The empirical application to S\&P 500 volatility forecasting evaluates the framework on three distinct base models: AR(1), ETS, and a modern LSTM neural network. The framework reduces cross origin incoherence by over 97\% across all models. Notably, enforcing coherence does not necessitate a sacrifice in point accuracy. For both the ETS and LSTM models, the Bayesian reconciled forecasts \emph{improve} the root mean squared error (RMSE) compared to the raw forecasts, suggesting that the latent state model acts as a powerful regularizer. Furthermore, the framework delivers fully calibrated probabilistic forecasts, with empirical coverage of the nominal 90\% intervals tightly calibrated between 93\% and 94\%, and highly competitive CRPS values. A Value at Risk (VaR) backtest confirms the practical utility of this probabilistic foundation, yielding a safer risk profile (4 breaches) compared to overconfident raw forecasts (6 breaches).

The approach is model agnostic and operates on already generated forecasts as a post processing layer, providing diagnostic information without requiring access to model internals. As demonstrated in the empirical application, the estimated horizon specific bias and reliability parameters act as an ``X-ray'' into the forecasting system. The framework successfully identified a massive negative bias at horizon 4 in the LSTM model and systematic over prediction at longer horizons in the ETS model, providing practitioners with actionable intelligence that deterministic stabilizers cannot offer.

Despite its strengths, the framework presents directions for future research. The current formulation assumes access to direct multi step forecasts and time invariant reliability. Extending the framework to accommodate recursively generated forecasts or time varying \(\tau_{h,t}^2\) would broaden its applicability in the presence of regime shifts. Also, while the Gaussian observation model performs well for moderately volatile series, extending the formulation to skewed or heavy tailed distributions would better capture the extreme right tail of strictly positive variables like volatility.

For large scale applications with extended forecast horizons, the computational cost of full MCMC inference could be reduced using variational Bayes approximations or state space methods such as the Kalman filter and smoother \citep{durbin2012}. More broadly, the proposed formulation could be extended to high dimensional systems and mixed frequency data. Taken together, the results demonstrate the feasibility of using Bayesian latent state inference to reconcile overlapping multi step forecasts. This post processing formulation complements existing work on forecast reconciliation and deterministic stabilizers by focusing on the construction of a coherent latent trajectory that simultaneously quantifies uncertainty and recovers horizon specific reliability, leaving the underlying forecasting model unchanged.

\section*{Funding Statement}
Ricardo Ehlers acknowledges financial support from the S\~ao Paulo Research Foundation (FAPESP), under grant 2025/06569-2, and from the National Council for Scientific and Technological Development (CNPq), under grant 409630/2024-1. Oluwatosin Bamidele Fajana acknowledges financial support from the Coordination for the Improvement of Higher Education Personnel (CAPES).

\section*{Data and Code Availability}
Complete replication code, including the Stan implementation of the Bayesian reconciliation model and all empirical experiments reported in this paper, is available at the \href{https://github.com/Fajtos/forecast-reconciliation/tree/main/Bayesian_Forecast_Reconciliation}{GitHub Repository}.

\section*{Declaration of Competing Interest}
The authors declare that they have no known competing financial interests or personal relationships that could have appeared to influence the work reported in this paper.\newpage

\bibliographystyle{plainnat}
\bibliography{references}

\end{document}